\documentclass[a4paper,11pt]{article}
\pdfoutput=1
\usepackage{graphicx}
\usepackage{amsmath}
\usepackage{amssymb}
\usepackage{latexsym}
\usepackage[colorlinks]{hyperref}
\usepackage{color}
\usepackage{cite}
\usepackage{makeidx}
\usepackage{float}
\newcommand{\bra}{\begin{array}}
\newcommand{\era}{\end{array}}
\newcommand{\beq}{\begin{equation}}
\newcommand{\eeq}{\end{equation}}
\newcommand{\bqr}{\begin{eqnarray}}
\newcommand{\eqr}{\end{eqnarray}}

\def\BC{\bb C}
\def\_\BC{\bbi C}

\def\( {\left(}
   \def\) {\right)}
\def\[ {\left[}
\def\] {\right]}
\def\no2 {{\textstyle{n\over 2}}}

\newcommand{\noi}{\noindent}

\begin{document}
\begin{titlepage}
\setcounter{page}{1}
\renewcommand{\thefootnote}{\fnsymbol{footnote}}

\begin{flushright}
%ucd-tpg 10-01\\
%arXiv:yymm.xxxx
\end{flushright}

\vspace{5mm}
\begin{center}

{\Large \bf {Transmission and Goos-H\"anchen like Shifts  through a
Graphene Double Barrier in an Inhomogeneous Magnetic Field}}

\vspace{5mm}
{\bf Miloud Mekkaoui}$^{a}$,
 {\bf Ahmed Jellal\footnote{\sf ajellal@ictp.it --
a.jellal@ucd.ac.ma}}$^{a,b}$ and {\bf Hocine Bahlouli}$^{b,c}$

\vspace{5mm}

{$^{a}$\em Theoretical Physics Group,  %Department of Physics,
Faculty of Sciences, Choua\"ib Doukkali University},\\
{\em PO Box 20, 24000 El Jadida, Morocco}

{$^b$\em Saudi Center for Theoretical Physics, Dhahran, Saudi Arabia}

{$^c$\em Physics Department,  King Fahd University
of Petroleum $\&$ Minerals,\\
Dhahran 31261, Saudi Arabia}

%\vspace{30mm}

\vspace{3cm}

\begin{abstract}
We studied the transport properties of electrons in graphene as they are scattered
by a double barrier potential in the presence of an inhomogeneous magnetic field.
We computed the transmission coefficient and Goos-H\"anchen like shifts for our system and
noticed that transmission is not allowed for certain range of energies.
In particular, we found that, in contrast to the electrostatic barriers, the magnetic barriers
are able to confine Dirac fermions. We also established some correlation between the
electronic transmission properties of Dirac fermions with the Goos-H\"anchen like shifts, as reflected in
the numerical data.
\vspace{3cm}

\end{abstract}
\end{center}


\end{titlepage}

%\newpage

%%%%%%%%%%%%%%%%%%%%%%%%%%%%%%%%%%%%%%%%%%%%%%%
\section{ Introduction}
%%%%%%%%%%%%%%%%%%%%%%%%%%%%%%%%%%%%%%%%%%%%%%%

Graphene, a planar arrangement of carbon atoms on a honeycomb
lattice, is a unique realization of a two dimensional electronic
system. Due to its excellent carrier transport properties, graphene
has a great potential for nano-electronic applications.
Among the peculiar electronic properties of this 2D-material is its
unusual quantum Hall effect \cite{Novoselov3}.
Graphene is also a transparent conductor \cite{Nair4} whose carriers
are massless and chiral relativistic fermions governed by a Dirac-like equation leading to
many fascinating physical properties of graphene, such as Klein
tunneling \cite{Katsnelson5,Yuanbo6}. However, as appealing as the
Klein tunneling may sound from the fundamental research point of
view, its presence in graphene is unwanted when it comes to
applications of graphene because space confinements of the
carriers is of great importance in nanoelectronic applications.
In addition, the ability to control electronic properties of a material
by an externally applied voltage is at the heart of modern
electronics \cite{Novoselov1,Yuanbo2}.

 The inability to confine electrons using an
electrostatic potential barrier severely limited the applicability
of graphene based devices. However, it came as a big relief when it was
pointed out that well localized magnetic field dubbed as magnetic barrier
can confine massless Dirac fermions in graphene \cite{martino}. Later on,
snake states, trajectories of charge carriers curving back and forth along interfaces,
were proven to play an important role and were studied experimentally \cite{Ye7,Lawton8},
mainly motivated by the quest for electrical rectification. The
inhomogeneous magnetic field case in graphene was analyzed in
\cite{martino}. Theoretically, electron waveguides, in
graphene subject to a suitable inhomogeneous magnetic field, were
considered in \cite{Ghosh9}. One of the interesting features of such
inhomogeneous magnetic field profile is that it can bind
electrons, contrary to the usual potential step. Such a step
magnetic field will indeed result in electron states that are
bound to the $B_{j}$-field step and are able to move only in one direction,
along the step.

 During the past few years there was substantial progress in studying electron
transport properties in graphene, among these developments we cite the quantum version of
the Goos-H\"anchen effect originating from the reflection of particles from
interfaces. Many works on various graphene-based nanostructures,
including single barrier \cite{Chen15}, double barrier \cite{Song16,
Jellal12} and superlattices \cite{Chen18}, showed that the
Goos-H\"anchen like (GHL) shifts can be enhanced by the transmission
resonances and controlled by varying the electrostatic potential
and induced gap \cite{Chen15}. Similar to the situation in semiconductors,
the GHL shifts in graphene can also be modulated by electric and
magnetic barriers \cite{Sharma19}, and atomic optics \cite{Huang13}.
It has been reported that the GHL shifts have a major effect on
the group velocity of quasiparticles along interfaces of graphene
p-n junctions \cite{Beenakker,Zhao11}. 

Very recently, the GHL
shifts for Dirac fermions in graphene scattered by double barrier
structures have been studied in \cite{Jellal12}. Moreover, in \cite{Jellal15}
we have explored the zero, positive and negative quantum GHL shifts of the transmitted
Dirac carriers in graphene through a potential barrier with vertical magnetic field. Numerical results
show that only one energy position at the zero GHL shift exists and is highly dependent on the $y$-directional
wave vector, the energy gap, the magnetic field and the potential. The positive and negative GHL
shifts happen when the incident energy is more and less than the energy position at the zero GHL shift,
respectively. In addition, we found that there are two values of potential at the zero GHL shifts, where a
potential window can always keep the positive GHL shifts. These results may be useful in designing a
graphene-based valley or spin splitter as well as manipulating the electrons and holes in graphene 
nanostructure. 

Motivated by different developments on the subject and in particular as a follow up on our
recent works \cite{Jellal12, Jellal15},
we investigate the GHL shifts in a gaped graphene system in the presence of
an inhomogeneous magnetic field and a double barrier potential. We
separate our system into three regions and determine the
solutions of the energy spectrum in each region.
Matching the wave functions at both interfaces, we then calculate the transmission
coefficient as well as the GHL shifts. To allow a better understanding of our
results, we study the transmission coefficient as well as the GHL shifts
while varying different physical parameters that characterize our system.

This paper is organized as follows. In section 2, we formulate our
system Hamiltonian describing particles scattered in graphene by a
double barrier potential in the presence of an inhomogeneous
magnetic field. We then obtain the solutions of the energy spectrum
corresponding to each region in terms of different physical
parameters and analyze the energy conservation law. In section 3, the scattering
problem for Dirac fermions will be solved using continuity at the boundary,
which will enable us to calculate the transmission coefficient and corresponding
phase. The condition for full reflection are then obtained for certain incidence angles
$\phi_{1}$. In section 4, we study the
GHL shifts and transmission coefficient as well as discuss our main results.
We present our main conclusions in the final section.

%%%%%%%%%%%%%%%%%%%%%%%%%%%%%%%%%%%%%%%%%%%%%%%%%%
\section{Theoretical model }
\begin{figure}[h]
  \centering
 \includegraphics[width=9cm, height=4cm ]{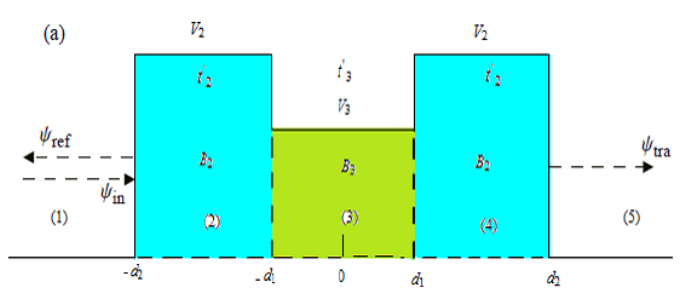}\ \ \ \
 \includegraphics[width=7cm, height=4cm ]{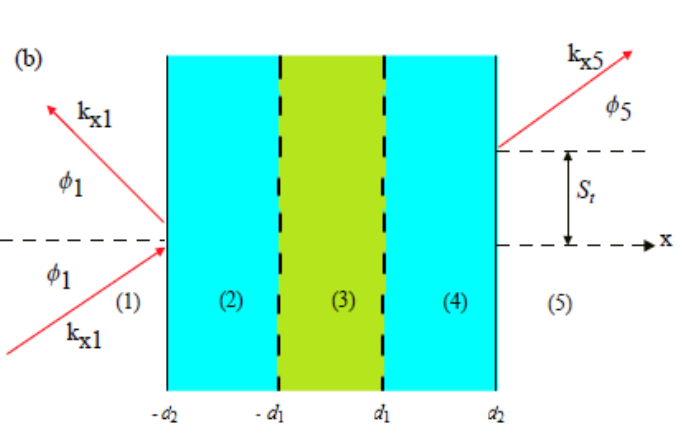}\\
  \caption{\sf Schematic diagram for Dirac fermions in an
inhomogeneous magnetic field and passing through a graphene double barrier,
with height $V_{2}$ in the region $d_{1}<\mid x\mid<d_{2}$ and
height $V_{3}$ in the region $\mid x \mid <d_{1}$. (a) the dashed lines show smooth electric potentials
having error function distributions. (b) describes the incident,
reflected, and transmitted electron beams with a lateral shift
$S_{t}$. }\label{fig1}
\end{figure}

%%%%%%%%%%%%%%%%%%%%%%%%%%%%%%%%%%%%%%%%%%%%%%%
We consider a system of massless Dirac fermions moving through a strip of
graphene and subject to a potential, which has the form shown
in the Figure \ref{fig1}. The system contains five regions denoted by the index
$j=1,2\cdots,5$. The left region ($j=1$) describes the incident
electron beam with energy $E=v_{F}\epsilon$ and incident
angle $\phi_{1}$ where $v_{F}$ is the Fermi velocity. The far
right region ($j=5$) describes the transmitted electron beam with
a lateral shift $S_{t}$ and angle $\phi_{5}$ but in the presence
of an inhomogeneous magnetic field. We introduce in the intermediate regions
$j=2,4$ and middle region $j=3$ two different magnetic fields
$B_{2}$ and $B_{3}$, respectively, such as
\begin{equation}\label{eq04}
B_{j}(x)=
\left\{%
\begin{array}{ll}
    B_{2}, & \hbox{$d_{1}<\mid x\mid<d_{2}$} \\
     B_{3}, & \hbox{$\mid x\mid<d_{1}$} \\
    0, & \hbox{otherwise.} \\
\end{array}%
\right.
\end{equation}
In the present study, we consider the  system in an inhomogeneous
magnetic field given by the configuration \eqref{eq04} in addition to the
presence of an energy gap $t_{j}^{'}$ in the regions 2, 3 and
4 defined by
\begin{equation}
t^{'}_{j}=
\left\{%
\begin{array}{ll}
    t^{'}_{2}, & \hbox{$d_{1}<\mid x\mid<d_{2}$} \\
     t^{'}_{3}, & \hbox{$\mid x\mid<d_{1}$} \\
    0, & \hbox{otherwise.} \\
\end{array}%
\right.
\end{equation}

\noi In order to study the scattering of Dirac fermions in graphene by
the above double barrier structure we first choose the following
potential configuration
\begin{equation}
V_{j}(x)=
\left\{%
\begin{array}{ll}
    V_{2}, & \hbox{$d_{1}<\mid x\mid<d_{2}$} \\
     V_{3}, & \hbox{$\mid x\mid<d_{1}$} \\
    0, & \hbox{otherwise} \\
\end{array}%
\right.
\end{equation}
where $j$ labels the five regions indicated schematically in
Figure \ref{fig1} that shows the space configuration of the
potential profile. The Hamiltonian for one-pseudospin component
in the $j$-th region can be written as
\begin{equation}\label{eq 4}
H_{j}=v_{F} {\boldsymbol{\sigma}}\cdot {\boldsymbol{\pi}}+
V_{j}(x)\mathbb{I}_{2}+t^{'}_{j}\sigma_{z}\Theta\left(d_{2}^{2}-x^{2}\right)
\end{equation}
where $\Theta$ is the Heaviside step function, $\pi=p+eA_{j}/c$
is the two-component kinetic momentum with the canonical momentum
$p=-i\hbar(\partial_{x},
\partial_{y})^{T}$, $ {\boldsymbol{\sigma}}= (\sigma_{x}, \sigma_{y})$ and $\sigma_{z}$  are
the usual Pauli matrices, $\mathbb{I}_{2}$ is the $2 \times 2$ unit matrix.
Choosing the Landau gauge we select the vector potential
$\textbf{A} = (0, Ay,  0)^{T}$ that creates the inhomogeneous magnetic field defined by \eqref{eq04},
imposing the continuity of this  vector potential at the boundaries of each region requires that
\begin{equation}
\qquad A_{y}(x)=A_{j}=\frac{c}{e}\times\left\{%
\begin{array}{ll}
    \frac{1}{l_{B_{2}}^{2}}(d_{1}-d_{2})-\frac{1}{l_{B_{3}}^{2}}d_{1}, & \hbox{$x<-d_{2}$} \\
     \frac{1}{l_{B_{2}}^{2}}x+(\frac{1}{l_{B_{2}}^{2}}-\frac{1}{l_{B_{3}}^{2}})d_{1}, & \hbox{$-d_{2}\leq x\leq -d_{1}$} \\
    \frac{1}{l_{B_{3}}^{2}}x, & \hbox{$\mid x\mid<d_{1}$} \\
   \frac{1}{l_{B_{2}}^{2}}x-(\frac{1}{l_{B_{2}}^{2}}-\frac{1}{l_{B_{3}}^{2}})d_{1}, & \hbox{$d_{1}\leq x\leq d_{2}$} \\
     \frac{1}{l_{B_{2}}^{2}}(d_{2}-d_{1})+\frac{1}{l_{B_{3}}^{2}}d_{1}, & \hbox{$x\geq d_{2}$} \\
\end{array}%
\right.
\end{equation}
where the local magnetic length is defined by $l_{B_{j}}=\sqrt{c/e B_{j}}$ in our selected
system of units $(\hbar= 1)$.

%%%%%%%%%%%%%%%%%%%%%%%%%%%%%%%%%%%%%%%%%%%%%%%%%%%%%%%%%%%%%%%%%%%%%%%%%%%%%%%%%%%%%%%%%
%\subsection{Energy spectrum solutions in the incident and transmission regions }
%%%%%%%%%%%%%%%%%%%%%%%%%%%%%%%%%%%%%%%%%%%%%%%%%%%%%%%%%%%%%%%%%%%%%%%%%%%%%%%%%%%%%

The eigenvalues and eigenspinors of %the Hamiltonian
$H_{j}$ in regions 1 and 5 are generated by the Dirac Hamiltonian
\begin{equation}
H_{j}=\left(%
\begin{array}{cc}
  0 & \upsilon_{F}\left[p_{xj}-i\left(p_{y}+\frac{c}{e}A_{j}\right)\right] \\
  \upsilon_{F}\left[p_{xj}+i\left(p_{y}+\frac{e}{c}A_{j}\right)\right] & 0 \\
\end{array}%
\right)
\end{equation}
and the time independent Dirac equation for the spinor $\psi_{j}(x,
y)= (\varphi_{j}^{+}, \varphi_{j}^{-})^{T}$ associated with energy
$E=\upsilon_{F}\epsilon$ is given by
\begin{equation}
H_{j}\left(%
\begin{array}{c}
  \varphi_{j}^{+} \\
  \varphi_{j}^{-} \\
\end{array}%
\right)=\epsilon\left(%
\begin{array}{c}
  \varphi_{j}^{+} \\
  \varphi_{j}^{-} \\
\end{array}%
\right)
\end{equation}
which can be written as two linear differential
equations of the form
\bqr
     &&\left[p_{xj}-i\left(p_{y}+\frac{e}{c}A_{j}\right)\right]\varphi_{j}^{-}=\epsilon\varphi_{j}^{+}\\
     &&\left[p_{xj}+i\left(p_{y}+\frac{e}{c}A_{j}\right)\right]\varphi_{j}^{+}=\epsilon\varphi_{j}^{-}.
\eqr
The corresponding energy eigenvalues read as
\begin{equation}
\epsilon=s_{j} \sqrt{p_{xj}^{2}+\left(p_{y}+\frac{e}{c}A_{j}\right)}
\end{equation}
where the symbol $s_{j}=\mbox{sign}(\epsilon)$ and
\begin{equation}
p_{xj}=\sqrt{\epsilon ^{2}-\left(p_{y}+\frac{e}{c}A_{j}\right)^{2}}.
\end{equation}
with incoming momentum ${\boldsymbol{p_{j}}}=(p_{xj}, p_{y})$ and position
${\boldsymbol{r}}=(x, y)$. The incoming wave function takes the form
\begin{equation}
\psi_{in}=\frac{1}{\sqrt{2}}\left(
\begin{array}{c}
1 \\
 z_{p_{xj}}\end{array}\right)e^{\textbf{\emph{i}}{\boldsymbol{p_{j}}}\cdot{\boldsymbol{r}}}
\end{equation}
and $z_{p_{xj}}$ is given by
\begin{equation}
z_{p_{xj}}=z_{j}=s_{j}\frac{p_{xj}
+i(p_{y}+\frac{e}{c}A_{j})}{\sqrt{(p_{xj})^{2}
+(p_{y}+\frac{e}{c}A_{j})^{2}}}=s_{j}
e^{\textbf{\emph{i}}\phi_{j}}
\end{equation}
where  $s_{0}=\mbox{sgn}(\epsilon)$ and  $\phi_{j}=\arctan\left(\frac{p_{y}-\frac{e}{c}A_{j}}{p_{xj}}\right)$ is the angle that
the incident electrons make with the {$x$-direction}, $p_{x1}$ and
$p_{y}$ are the $x$ and $y$-components of the electron wave vector, respectively. The eigenspinors read as
\bqr
&&\psi_{j}^{+}=\frac{1}{\sqrt{2}}\left(
\begin{array}{c}
1 \\
 z_{j}\end{array}\right)e^{\textbf{\emph{i}}(p_{xj} x +p_{y} y)}
\\
&& \psi_{j}^{-}=\frac{1}{\sqrt{2}}\left(
\begin{array}{c}
1 \\
 -z^{*}_{j}\end{array}\right)e^{\textbf{\emph{i}}(-p_{xj} x +p_{y} y)}.
\eqr
%It is straightforward to solve the tunneling problem for Dirac
%fermions. We assume that the incident  wave propagates at an angle
%$\phi_{1}$ with respect to the {$x$-direction} and then use  the
%components, of the Dirac spinor $\varphi_{j}^{+}$ and
%$\varphi_{j}^{-}$, in each region, in the following form :\\
To be much more accurate, we give the solutions of the energy spectrum for
each region. Then 
in region 1 ( $x<-d_{2}$ ), we have
\bqr
&& \epsilon=
\sqrt{p_{x1}^{2}+\left[p_{y}+\frac{1}{l_{B_{2}}^{2}} (d_{1}-d_{2})-\frac{1}{l_{B_{3}}^{2}}d_{1}\right]^{2}}
\\
&&
\psi_{1}=\frac{1}{\sqrt{2}}\left(
\begin{array}{c}
1 \\
 z_{1}\end{array}\right)e^{\textbf{\emph{i}}(p_{x1} x +p_{y} y)}+r\frac{1}{\sqrt{2}}\left(
\begin{array}{c}
1 \\
 -z^{*}_{1}\end{array}\right)e^{\textbf{\emph{i}}(-p_{1x} x +p_{y}
 y)}
\\
&&
z_{1}=s_{1}\frac{p_{x1}
+i\left[p_{y}+\frac{1}{l_{B_{2}}^{2}}(d_{1}-d_{2})-\frac{1}{l_{B_{3}}^{2}}d_{1}\right]}{\sqrt{p_{x1}^{2}
+\left[p_{y}+\frac{1}{l_{B_{2}}^{2}}(d_{1}-d_{2})-\frac{1}{l_{B_{3}}^{2}}d_{1}\right]^{2}}}
\eqr
and in region 5 $( x > d_{2})$, the solution is
\bqr
&& \epsilon=\sqrt{p_{x5}^{2}+\left[p_{y}+\frac{1}{l_{B_{2}}^{2}}(d_{2}-d_{1})+\frac{1}{l_{B_{3}}^{2}}d_{1}\right]^{2}}
\\
&&
\Psi_{5}=\frac{1}{\sqrt{2}}t\left(
\begin{array}{c}
1 \\
 z_{5}\end{array}\right)e^{\textbf{\emph{i}}(p_{x5} x +p_{y} y)}
\\
&&
z_{5}=s_{5}\frac{p_{x5}
+i\left[p_{y}+\frac{1}{l_{B_{2}}^{2}}(d_{2}-d_{1})+\frac{1}{l_{B_{3}}^{2}}d_{1}\right]}{\sqrt{p_{x1}^{2}
+\left[p_{y}+\frac{1}{l_{B_{2}}^{2}}(d_{2}-d_{1})+\frac{1}{l_{B_{3}}^{2}}d_{1}\right]^{2}}}.
\eqr

%%%%%%%%%%%%%%%%%%%%%%%%%%%%%%%%%%%%%%%%%%%%%%%
%\subsection{Energy spectrum solutions in regions $|x|\leq d_{2}$  }
%%%%%%%%%%%%%%%%%%%%%%%%%%%%%%%%%%%%%%%%%%%%%%%
For the system under consideration, we can write the
Hamiltonian corresponding to regions (2), (3) and (4) in
matrix form as
\begin{equation}\label{eq 20}
H_{j}=v_{F}\left(%
\begin{array}{cc}
  \frac{V_{j}}{v_{F}}+\frac{t^{'}_{j}}{v_{F}} & -i\frac{\sqrt{2}}
  {l_{B_{j}}} \left[\frac{l_{B_{j}}}{\sqrt{2}} \left(\partial_{xj}-i\partial_{y}+\frac{e}{c}A_{j} \right)\right]\\
 i\frac{\sqrt{2}}{l_{B_{j}}} \left[\frac{l_{B_{j}}}{\sqrt{2}} \left(-\partial_{xj}-i\partial_{y}
 +\frac{e}{c}A_{j} \right)\right]  &  \frac{V_{j}}{v_{F}}-\frac{t^{'}_{j}}{v_{F}}\\
\end{array}%
\right).
\end{equation}
Note that the energy gap $t^{'}_{j}$ is equivalent to a mass term,
this will lead to interesting consequences on the physical properties of such
system. We determine the eigenvalues and eigenspinors of the
corresponding Hamiltonian $H$ by solving the time independent equation for the spinor
$\psi_{j}(x, y)=(\psi_{j}^{+}, \psi_{j}^{-})^{T}$. Since the
transverse momentum $p_{y}$ is conserved, we can then write the wave function as
%\begin{equation}
$\psi_{j}(x, y)=e^{ip_{y}y} \varphi_{j}(x)$,
%\end{equation}
with $\varphi_{j}(x)=
(\varphi_{j}^+, \varphi_{j}^-)^{T}$, and energy
$E=\upsilon_{F}\epsilon$, which lead to
\begin{equation}\label{eq 23}
H_{j}\left(%
\begin{array}{c}
  \varphi_{j}^+ \\
  \varphi_{j}^-\\
\end{array}%
\right)=\epsilon\left(%
\begin{array}{c}
  \varphi_{j}^+\\
  \varphi_{j}^-\\
\end{array}%
\right).
\end{equation}
At this stage, it is convenient to introduce the concepts of annihilation and
creation operators in order to ease the diagonalization of our Hamiltonian. They can be defined by
\begin{eqnarray}
a_{j}=\frac{l_{B_{j}}}{\sqrt{2}} \left(\partial_{xj}+k_{y}+\frac{e}{c}A_{j}\right),
\qquad
a_{j}^{\dagger}=\frac{l_{B_{j}}}{\sqrt{2}}\left(-\partial_{xj}+k_{y}+\frac{e}{c}A_{j}\right)
\end{eqnarray}
and obey the canonical commutation relations $\left[a_{j},
a_{k}^{\dagger}\right]=\delta_{j,k}$. Rescaling our energies
$t^{'}_{j}=\upsilon_{F}\mu_{j}$ and $V_{j}=\upsilon_{F}v_{j}$, then
 \eqref{eq 23} can be written in terms of $a_{j}$ and
$a^{\dagger}_{j}$ as
\begin{equation}
 \left(%
\begin{array}{cc}
  v_{j}+\mu_{j} & -i\frac{\sqrt{2}}{l_{B_{j}}}a_{j} \\
  +i\frac{\sqrt{2}}{l_{B_{j}}}a_{j}^{\dagger}  &  v_{j}-\mu_{j} \\
\end{array}%
\right)\left(%
\begin{array}{c}
  \varphi_{j}^+ \\
  \varphi_{j}^- \\
\end{array}%
\right)=\epsilon\left(%
\begin{array}{c}
  \varphi_{j}^+ \\
  \varphi_{j}^- \\
\end{array}%
\right)
\end{equation}
giving rise to the two relations between spinor
components
\bqr
&&  (v_{j}+\mu_{j})\varphi_{j}^{+}-i\frac{\sqrt{2}}{l_{B_{j}}}a_{j}\varphi_{j}^-=\epsilon\varphi_{j}^+ \label{eq 25}\\
 && i\frac{\sqrt{2}}{l_{B_{j}}}a_{j}^{\dagger}\varphi_{j}^+ +
  (v_{j}-\mu_{j})\varphi_{j}^{-}=\epsilon\varphi_{j}^- \label{eq 26}.
\eqr
Injecting \eqref{eq 26} in \eqref{eq 25}, we obtain a second order
differential equation for $\varphi_{j}^{+}$
\begin{equation}
\left[\left(\epsilon-v_{j}\right)^{2}-\mu^{2}_{j}\right]\varphi_{j}^{+}=\frac{2}{l_{B_{j}}^{2}}a_{j}
a_{j}^{\dagger}\varphi_{j}^{+}
\end{equation}
which shows clearly that $\varphi_{j}^{+}$ is an eigenstate of the number
operator $\widehat{N_{j}}=a_{j}^{\dagger}a_{j}$ and therefore we
identify $\varphi_{j}^{+}$ to be eigenstates of the harmonic
oscillator $|n_{j}-1\rangle$, namely
\begin{equation}
 \varphi_{j}^{+} \sim \mid n_{j}-1\rangle
\end{equation}
which is equivalent to stating
 \begin{equation}
 \left[\left(\epsilon-v_{j}\right)^{2}-\mu^{2}_{j}\right] \mid
n_{j}-1\rangle=\frac{2}{l_{B_{j}}^{2}}n_{j}\mid n_{j}-1\rangle
\end{equation}
and the energy spectrum can be defined by
\begin{equation}
\epsilon-v_{j}=s_{j}
\epsilon_{n_{j}}=s_{j}\frac{1}{l_{B_{j}}}\sqrt{\left(\mu_{j}
l_{B_{j}}\right)^{2}+2n_{j}}
\end{equation}
where we have set $\epsilon_{n_{j}}=s_{j}\left(\epsilon-v_{j}\right)$ and
$s_{j}=\mbox{sign}\left(\epsilon_{n_{j}}-v_{j}\right)$ corresponding to positive and
negative energy solutions.
The second spinor component now reads as 
\begin{equation}
\varphi_{j}^{-}=s_{j}i\sqrt{\frac{\epsilon_{n_{j}}l_{B_{j}}-s_{j}
\mu_{j} l_{B_{j}}}{\epsilon_{n_{j}}l_{B_{j}}+s_{j} \mu_{j}
l_{B_{j}}}} \mid n_{j}\rangle.
\end{equation}
After normalization we arrive at the expression for the positive and negative energy
eigenstates
\begin{equation}
\varphi_{j}=\frac{1}{\sqrt{2}}\left(%
\begin{array}{c}
  \sqrt{\frac{\epsilon_{n_{j}}l_{B_{j}}+s_{j} \mu_{j} l_{B_{j}}}{\epsilon_{n_{j}}l_{B_{j}}}} \mid n_{j}-1\rangle \\
  s_{j} i\sqrt{\frac{\epsilon_{n_{j}}l_{B_{j}}-s_{j} \mu_{j} l_{B_{j}}}{\epsilon_{n_{j}}l_{B_{j}}}} \mid n_{j}\rangle \\
\end{array}%
\right).
\end{equation}
Introducing the parabolic cylinder functions
$D_{n_{j}}(x)=2^{-\frac{n_{j}}{2}}e^{-\frac{x^{2}}{4}}H_{n_{j}} \left(\frac{x}{\sqrt{2}}\right)$ to express
the solution in regions 2, 3 and 4 as 
\begin{equation}
\psi_{j}^{\pm}(x, y)=\frac{1}{\sqrt{2}}\left(%
\begin{array}{c}
 \sqrt{\frac{\epsilon_{n_{j}}l_{B_{j}}+s_{j} \mu_{j} l_{B_{j}}}{\epsilon_{n_{j}}l_{B_{j}}}}
 D_{\left(\left(\epsilon_{n_{j}}l_{B_{j}}\right)^{2}-\left(\mu_{j} l_{B_{j}}\right)^{2}
 \right)/2-1}
 \left(\pm \sqrt{2}\left(\frac{x}{l_{B_{j}}}+k_{y}l_{B_{j}}\right)\right) \\
  \frac{\pm i s_{j}\sqrt{2}}{\sqrt{\epsilon_{n_{j}}l_{B_{j}}\left(\epsilon_{n_{j}}l_{B_{j}}+s_{j} \mu_{j} l_{B_{j}}\right)}}
  D_{\left(\left(\epsilon_{n_{j}}l_{B_{j}}\right)^{2}-\left(\mu_{j} l_{B_{j}}\right)^{2}\right)/2}
  \left(\pm \sqrt{2}\left(\frac{x}{l_{B_{j}}}+k_{y}l_{B_{j}}\right)\right) \\
\end{array}%
\right)e^{ik_{y}y}.
\end{equation}
In summary the solutions of the energy spectrum in the barrier $( -d_{2}\leq x \leq -d_{1})$ (region 2) are
\bqr
&&\epsilon_2=v_{2}+s_{2}\frac{1}{l_{B_{2}}}\sqrt{(\mu_{2}
l_{B_{2}})^{2}+2n_{2}}
\\
&&
\psi_{2}(x,y)=a_{2}\psi_{2}^{+}+b_{2}\psi_{2}^{-}
\eqr
while in region 3 $( |x|\leq d_{1})$ read as
\bqr
&& \epsilon_3=v_{3}+s_{3}\frac{1}{l_{B_{3}}}\sqrt{(\mu_{3}
l_{B_{3}})^{2}+2n_{3}}
\\
&&
\psi_{3}(x,y)=a_{3}\psi_{3}^{+}+b_{3}\psi_{3}^{-}
\eqr
and finally %the eigenspinor
in region 4 $( d_{1}\leq x \leq
d_{2})$ it can be expressed as
\bqr
&& \epsilon_4=v_{2}+s_{4}\frac{1}{l_{B_{2}}}\sqrt{(\mu_{2}
l_{B_{2}})^{2}+2n_{4}}
\\
&& \psi_{4}(x,y)=a_{4}\psi_{4}^{+}+b_{4}\psi_{4}^{-}
\eqr
where the parameters $a_j$ and $b_j$, with $(j=2,3,4)$, are normalization constants.

 Recall that,
from the above analysis, we ended up with different energy spectra
$\epsilon_2$, $\epsilon_3$ and $\epsilon_4$, which
are obtained in terms of system parameters
and quantum numbers in each regions. On the other hand, energy conservation requires that
\begin{equation}
\epsilon=\epsilon_2=\epsilon_3=\epsilon_4
\end{equation}
and by replacing the energies by their expressions, it is easy to observe that the
allowed energy values should satisfy the relation
\begin{equation}
n_{2}=n_{4}=\frac{l_{B_{2}}^{2}}{2}\left[\left(v_{3}-v_{2}+s_{3}\sqrt{\mu^{2}_{3}+\frac{2n_{3}}{l_{B_{3}}^{2}}}\right)^{2}
-\mu^{2}_{2}
 \right].
\end{equation}

Having obtained all solutions of the energy spectrum, we will see how they can
be used to investigate different physical properties of our system. Specifically, we
evaluate the transmission and reflection amplitudes in terms of
different physical system parameters.

%%%%%%%%%%%%%%%%%%%%%%%%%%%%%%%%%%%%%%%%%%%%%%%%%%%%%%%
 \section{Transmission and phase shift}
%%%%%%%%%%%%%%%%%%%%%%%%%%%%%%%%%%%%%%%%%%%%%%%%%%%%%%%%

%\hspace{.33in}
Before determining explicitly the transmission coefficient and its associated
phase shift, we notice that total internal reflection will take
place only when $0 < \phi_{1} < \frac{\pi}{2}$, since the wave
incident from the right-hand and left-hand side of the normal surface
will behave differently \cite{Ghosh}. It is clear that the shift
in $p_{y}$ is due to our choice of gauge for the vector potential. We find it more
convenient to parameterize the momenta by
\bqr
&& p_{x1}=\epsilon\cos\phi_{1},\qquad
p_{y}=\epsilon\sin\phi_{1}+\frac{1}{l_{B_{2}}^{2}}(d_{2}-d_{1})+\frac{d_{1}}{l_{B_{3}}^{2}}
\\
&&
p_{x5}=\epsilon\cos\phi_{5},\qquad
p_{y}=\epsilon\sin\phi_{5}-\frac{1}{l_{B_{2}}^{2}}(d_{2}-d_{1})-\frac{d_{1}}{l_{B_{3}}^{2}}.
\eqr
It is clear that the refraction angles $\phi_{5}$ at the interfaces are
obtained by requiring conservation of the momentum $p_{y}$. This
leads to a simplified expression of these angles in terms of
$\phi_{1}$
\begin{equation}\label{eq 51}
\sin\phi_{5}=\sin\phi_{1}+\frac{2}{\epsilon
l_{B_{2}}^{2}}(d_{2}-d_{1})+\frac{2d_{1}}{\epsilon l_{B_{3}}^{2}}
\end{equation}
and therefore
we characterize our waves by introducing a critical angle
$\phi_{c}$
\begin{equation}
\phi_{c}=\sin^{-1}\left[1+2d_{1}\left(\frac{1}{\epsilon
l_{B_{2}}^{2}}-\frac{1}{\epsilon
l_{B_{3}}^{2}}\right)-\frac{2d_{2}}{\epsilon l_{B_{2}}^{2}}
\right].
\end{equation}
This tells us  that when the incident angle is less than $\phi_c$, the
modes become oscillating guided modes, while in the case when the
incident angle is more than $\phi_{c}$, we obtain decaying or
evanescent wave modes.

In the forthcoming analysis, we will be
interested in studying the situation where $\phi_{1} < \phi_{c}$.
To simplify our task and proceed further, let us choose the
interfaces separating regions as
\begin{equation}
a_{n_{j}}=\sqrt\frac{{\epsilon_{n_{j}}l_{B_{j}}+s_{j} \mu_{j}
l_{B_{j}}}}{{\epsilon_{n_{j}}l_{B_{j}}}},\qquad
b_{n_{j}}=\frac{s_{j}\sqrt{2}}{\sqrt{\epsilon_{n_{j}}l_{B_{j}}(\epsilon_{n_{j}}l_{B_{j}}+s_{j}
\mu_{j} l_{B_{j}})}}.
\end{equation}
We match  the wave functions at the boundaries
$(-d_{2},-d_{1},d_{1},d_{2})$ as required by the first order
nature of the Dirac equation. For this, we introduce the shorthand notations
% we end up with the following set of
%equations
\bqr \label{eq11} &&
\eta_{1n_{2}}^{\pm}=D_{\left(\left(\epsilon_{n_{2}}l_{B_{2}}\right)^{2}-\left(\mu_{2}
l_{B_{2}}\right)^{2}
 \right)/2-1}\left(\pm\sqrt{2}
\left(\frac{-d_{2}}{l_{B_{2}}}+k_{y}l_{B_{2}}\right)\right)  \\ %\nonumber
&& \xi_{1n_{2}}^{\pm}=D_{\left(\left(\epsilon_{n_{2}}l_{B_{2}}\right)^{2}-\left(\mu_{2}
l_{B_{2}}\right)^{2}\right)/2}
\left(\pm\sqrt{2}\left(\frac{-d_{2}}{l_{B_{2}}}+k_{y}l_{B_{2}}\right)\right)   %\nonumber
\eqr
the related symbols $\eta_{2n_{2}}^{\pm}$,
$\xi_{2n_{2}}^{\pm}$ follow by letting $-d_{2}\longrightarrow
-d_{1}$,
\bqr \label{eq12} &&
\eta_{1n_{3}}^{\pm}=D_{\left(\left(\epsilon_{n_{3}}l_{B_{3}}\right)^{2}-\left(\mu_{3}
l_{B_{3}}\right)^{2}
 \right)/2-1}\left(\pm\sqrt{2}
\left(\frac{-d_{1}}{l_{B_{3}}}+k_{y}l_{B_{3}}\right)\right)  \\
&& \xi_{1n_{3}}^{\pm}=D_{\left(\left(\epsilon_{n_{3}}l_{B_{3}}\right)^{2}-\left(\mu_{3}
l_{B_{3}}\right)^{2}\right)/2}
\left(\pm\sqrt{2}\left(\frac{-d_{1}}{l_{B_{3}}}+k_{y}l_{B_{3}}\right)\right) %\nonumber
\eqr
the related symbols $\eta_{2n_{3}}^{\pm}$,
$\xi_{2n_{3}}^{\pm}$ follow by letting $-d_{1}\longrightarrow
d_{1}$, %. Similarly,
 \bqr \label{eq12} &&
\eta_{1n_{4}}^{\pm}=D_{\left(\left(\epsilon_{n_{4}}l_{B_{2}}\right)^{2}-\left(\mu_{2}
l_{B_{2}}\right)^{2}
 \right)/2-1}\left(\pm\sqrt{2}
\left(\frac{d_{1}}{l_{B_{2}}}+k_{y}l_{B_{2}}\right)\right)  \\
&&
\xi_{1n_{4}}^{\pm}=D_{\left(\left(\epsilon_{n_{4}}l_{B_{2}}\right)^{2}-\left(\mu_{2}
l_{B_{2}}\right)^{2}\right)/2}
\left(\pm\sqrt{2}\left(\frac{d_{1}}{l_{B_{2}}}+k_{y}l_{B_{2}}\right)\right) %\nonumber
\eqr
the related symbols $\eta_{2n_{4}}^{\pm}$,
$\xi_{2n_{4}}^{\pm}$ follow by letting $d_{1}\longrightarrow
d_{2}$.  Now, requiring the continuity of the spinor wavefunctions
at each junction interface give rise to a set of equations which can be
expressed in terms of $2\times 2$ transfer matrices between different regions
%, $M_{jj+1}$
%where
\beq
\left(%
\begin{array}{c}
  a_{j} \\
  b_{j} \\
\end{array}%
\right)=M_{j j+1}\left(%
\begin{array}{c}
  a_{j+1} \\
  b_{j+1} \\
\end{array}%
\right)
\eeq
where $M_{j j+1}$ is a transfer matrix that couple the
wave function in the $j$-th region to the wave function in the $(j
+ 1)$-th region. Finally, we obtain the full transfer matrix over the
whole double barrier region, which can be expressed in an obvious notation
as
\begin{equation}
\left(%
\begin{array}{c}
  a_{1} \\
  b_{1} \\
\end{array}%
\right)=\prod_{j=1}^{4}M_{j j+1}\left(%
\begin{array}{c}
  a_{5} \\
  b_{5} \\
\end{array}%
\right)=M\left(%
\begin{array}{c}
  a_{5} \\
  b_{5} \\
\end{array}%
\right).
\end{equation}
The total transfer matrix $M=M_{12}\cdot M_{2
3}\cdot M_{34}\cdot M_{45}$ is a transfer matrix that couple the
wave function in the incident region to the wave function in the transmission region.
It can be expressed explicitly as
% it is given by
\bqr
 M &=&\left(%
\begin{array}{cc}
  m_{11} & m_{12} \\
  m_{21} & m_{22} \\
\end{array}%
\right)
\\
M_{12} &=&\left(%
\begin{array}{cc}
   e^{-\textbf{\emph{i}}p_{x1} d_{2}} &e^{\textbf{\emph{i}}p_{x1} d_{2}} \\
  z_{1}e^{-\textbf{\emph{i}}p_{x1} d_{2}} & -z^{\ast}_{1} e^{\textbf{\emph{i}}p_{x1} d_{2}} \\
\end{array}%
\right)^{-1}\left(%
\begin{array}{cc}
  a_{n_{2}}\eta_{1n_{2}}^{+} & a_{n_{2}}\eta_{1n_{2}}^{-} \\
 ib_{n_{2}}\xi_{1n_{2}}^{+} & -ib_{n_{2}} \xi_{1n_{2}}^{-} \\
\end{array}%
\right)
\\
M_{23} &=&\left(%
\begin{array}{cc}
  a_{n_{2}}\eta_{2n_{2}}^{+} & a_{n_{2}}\eta_{2n_{2}}^{-} \\
  ib_{n_{2}}\xi_{2n_{2}}^{+}  & -i b_{n_{2}}\xi_{2n_{2}}^{-}\\
\end{array}%
\right)^{-1}\left(%
\begin{array}{cc}
  a_{n_{3}}\eta_{1n_{3}}^{+}  & a_{n_{3}}\eta_{1n_{3}}^{-} \\
  ib_{n_{3}}\xi_{1n_{3}}^{+}  & -ib_{n_{3}} \xi_{1n_{3}}^{-} \\
\end{array}%
\right)
\\
M_{34} &=&\left(%
\begin{array}{cc}
  a_{n_{3}}\eta_{2n_{3}}^{+}  & a_{n_{3}}\eta_{2n_{3}}^{-} \\
  ib_{n_{3}}\xi_{2n_{3}}^{+} & -i b_{n_{3}}\xi_{2n_{3}}^{-} \\
\end{array}%
\right)^{-1}\left(%
\begin{array}{cc}
  a_{n_{2}}\eta_{1n_{4}}^{+} & a_{n_{2}}\eta_{1n_{4}}^{-} \\
  ib_{n_{2}}\xi_{1n_{4}}^{+} & -ib_{n_{2}} \xi_{1n_{4}}^{-} \\
\end{array}%
\right)
\\
M_{45} &=& \left(%
\begin{array}{cc}
  a_{n_{2}}\eta_{2n_{4}}^{+}  & a_{n_{2}}\eta_{2n_{4}}^{-} \\
  ib_{n_{2}}\xi_{2n_{4}}^{+} & -i b_{n_{2}}\xi_{2n_{4}}^{-}  \\
\end{array}%
\right)^{-1}\left(%
\begin{array}{cc}
  e^{\textbf{\emph{i}}p_{x5} d_{2}} & e^{-\textbf{\emph{i}}p_{x5} d_{2}} \\
  z_{5} e^{\textbf{\emph{i}}p_{x5} d_{2}}  & -z_{5}^{\ast} e^{-\textbf{\emph{i}}p_{x5} d_{2}}  \\
\end{array}%
\right).
\eqr
We consider an electron propagating from left to right with energy
$\epsilon l_{B_{2}}$, then $r=b_{1}$ and $t=a_{5}$, $r$ and
$t$ being the reflection and transmission amplitudes,
respectively. We have assumed an incident wave from left
normalized to unit amplitude $a_{1} =1$ and $b_{5} =0$ is the null
amplitude due absence of left moving waves in transmission region.
This will give rise to the following relations
%the reflection and
%transmission amplitudes
\begin{equation}\label{eq 63}
 t=\frac{1}{m_{11}}, \qquad  r=\frac{m_{21}}{m_{11}}.
\end{equation}
This last formulation will be much more adequate in dealing with
periodic systems and applying Bloch theorem to find the
associated energy bands. The above expressions  can be
written as 
\beq\label{eq 62}
 t=\frac{1}{\left|m_{11}\right|} e^{i\varphi_{t}},\qquad r=\left|\frac{m_{21}}{m_{11}}\right| e^{i\varphi_{r}}
\eeq
where $\varphi_{t}$ and $\varphi_{r}$ refers to the phase of
the transmission and reflection amplitudes, respectively.
%and the subscript corresponds to the wavevector at central incidence angle.
After a lengthy but straightforward
algebra, we can show that $t$ in \eqref{eq 62} %the transmission coefficient
takes the form
\begin{eqnarray}
%\frac{1}{|m_{11}|} e^{i\varphi_{t}}
t=s_{3}a_{n_{2}}^{2}b_{n_{2}}^{2}a_{n_{3}}
b_{n_{3}}\lambda_{n_{2}}\lambda_{n_{3}}\lambda_{n_{4}}\frac{\Lambda^{+}\chi^{+}+\Lambda^{-}\chi^{-}+
i(\Lambda^{-}\chi^{+}-\Lambda^{+}\chi^{-})}{(\chi^{+})^{2}+(\chi^{-})^{2}}
\end{eqnarray}
where we have set
\begin{eqnarray*}
\Lambda^{+}&=&\left(1+(q_{1}^{+})^{2}-(q^{-}_{1})^{2}\right)\sin(d_{2}(p_{x1}+p_{x5}))-2q_{1}^{+}q_{1}^{-}\cos(d_{2}(p_{x1}+p_{x5}))\\
\Lambda^{-}&=&\left(1+(q_{1}^{+})^{2}-(q^{-}_{1})^{2}\right)\cos(d_{2}(p_{x1}+p_{x5}))+2q_{1}^{+}q_{1}^{-}\sin(d_{2}(p_{x1}+p_{x5}))\\
\chi^{+}&=&-a_{n_{2}}^{2}b_{n_{3}}^{2}\beta_{n_{3}}D-a_{n_{3}}^{2}b_{n_{2}}^{2}\alpha_{n_{3}}C+
s_{2}s_{3}a_{n_{2}}a_{n_{3}}b_{n_{2}}b_{n_{3}}(q_{1}^{+}B_{1}+q_{5}^{+}B_{2})\\
 &&-s_{3}a_{n_{2}}^{2}a_{n_{3}}b_{n_{2}}^{2}b_{n_{3}}A_{1}(q_{1}^{+}q_{5}^{-}-q_{5}^{+}q_{1}^{-})\\
\chi^{-}&=&-a_{n_{2}}^{2}b_{n_{3}}^{2}\beta_{n_{3}}E-a_{n_{3}}^{2}b_{n_{2}}^{2}\alpha_{n_{3}}F+
s_{2}s_{3}a_{n_{2}}a_{n_{3}}b_{n_{2}}b_{n_{3}}(q_{1}^{-}B_{1}+q_{5}^{-}B_{2})\\
&&
+s_{3}a_{n_{2}}^{2}a_{n_{3}}b_{n_{2}}^{2}b_{n_{3}}((q_{1}^{+}q_{5}^{+}-q_{1}^{-}q_{5}^{-})A_{1}+A_{2})\\
%\end{eqnarray*}
%\begin{eqnarray*}
A_{1}&=&\delta_{n_{2}}\delta_{n_{3}}\delta_{n_{4}}+\beta_{n_{2}}\gamma_{n_{3}}\alpha_{n_{4}}\\
A_{2}&=&\gamma_{n_{2}}\gamma_{n_{3}}\gamma_{n_{4}}+\alpha_{n_{2}}\delta_{n_{3}}\beta_{n_{4}}\\
B_{1}&=&b_{n_{2}}^{2}(\beta_{n_{2}}\gamma_{n_{3}}\gamma_{n_{4}}+\delta_{n_{2}}\delta_{n_{3}}\beta_{n_{4}})\\
B_{2}&=&-a_{n_{2}}^{2}(\alpha_{n_{2}}\delta_{n_{3}}\delta_{n_{4}}+\gamma_{n_{2}}\gamma_{n_{3}}\alpha_{n_{4}})\\
 C &=&q_{5}^{+}a_{n_{2}}\delta_{n_{4}}\left(s_{2}q_{1}^{-}b_{n_{2}}\beta_{n_{2}}+a_{n_{2}}\gamma_{n_{2}}\right)+s_{2}b_{n_{2}}q_{1}^{+}
 \beta_{n_{2}}(q_{5}^{-}a_{n_{2}}\alpha_{n_{4}}-s_{2}b_{n_{2}}\gamma_{n_{4}})\\
 D &=&q^{+}_{5}a_{n_{2}}\alpha_{n_{4}}(a_{n_{2}}\alpha_{n_{2}}-s_{2}b_{n_{2}}\delta_{n_{2}})+
 s_{2}q^{+}_{1}b_{n_{2}}\delta_{n_{2}}(q_{5}^{-}a_{n_{2}}\alpha_{n_{4}}-s_{2}b_{n_{2}}\gamma_{n_{4}})\\
E &=& (a_{n_{2}}\alpha_{n_{2}}-s_{2}b_{n_{2}}\delta_{n_{2}})(q_{5}^{-}a_{n_{2}}\alpha_{n_{4}}-
s_{2}b_{n_{2}}\gamma_{n_{4}})-s_{2}q^{+}_{5}a_{n_{2}}\alpha_{n_{4}}q^{+}_{1}b_{n_{2}}\delta_{n_{2}}\\
F &=& (s_{2}q_{1}^{-}b_{n_{2}}\beta_{n_{2}}+a_{n_{2}}\gamma_{n_{2}})(q_{5}^{-}a_{n_{2}}\alpha_{n_{4}}-s_{2}b_{n_{2}}\gamma_{n_{4}})
-s_{2}q_{5}^{+}a_{n_{2}}\delta_{n_{4}}b_{n_{2}}q_{1}^{+}
 \beta_{n_{2}}\\
%\end{eqnarray*}
%\begin{eqnarray*}
z_{j}&=&q^{+}_{j}+iq^{-}_{j}\\
\alpha_{n_{j}} &=&
\eta_{1n_{j}}^{-}\eta_{2n_{j}}^{+}-\eta_{1n_{j}}^{+}
 \eta_{2n_{j}}^{-}\\
\beta_{n_{j}} &=&
\xi_{1n_{j}}^{-}\xi_{2n_{j}}^{+}-\xi_{1n_{j}}^{+}\xi_{2n_{j}}^{-}\\
 \gamma_{n_{j}} &=&
\eta_{1n_{j}}^{-}\xi_{2n_{j}}^{+}+\eta_{1n_{j}}^{+}
 \xi_{2n_{j}}^{-}\\
 \delta_{n_{j}} &=&
\eta_{2n_{j}}^{-}\xi_{1n_{j}}^{+}+\eta_{2n_{j}}^{+}
 \xi_{1n_{j}}^{-}\\
 \lambda_{n_{j}} &=&
\eta_{2n_{j}}^{-}\xi_{2n_{j}}^{+}+\eta_{2n_{j}}^{+}
 \xi_{2n_{j}}^{-}.
\end{eqnarray*}
The phase shift can be expressed explicitly as
 \beq \label{eq phase}
 \varphi_{t}=\arctan\left[\frac{\Lambda^{-}\chi^{+}-\Lambda^{+}\chi^{-}}{\Lambda^{+}\chi^{+}+\Lambda^{-}\chi^{-}}\right]
\eeq
with the quantities
\begin{eqnarray}
&&\Lambda^{-}\chi^{+}-\Lambda^{+}\chi^{-}=2\cos\phi_{1}(\chi^{+}\cos(\phi_{1}+(p_{x1}+p_{x5}))-\chi^{-}\sin(\phi_{1}-(p_{x1}+p_{x5}))\\
&&\Lambda^{+}\chi^{+}+\Lambda^{-}\chi^{-}=2\cos\phi_{1}(\chi^{+}\sin(\phi_{1}-(p_{x1}+p_{x5}))+\chi^{-}\cos(\phi_{1}+(p_{x1}+p_{x5})).
\end{eqnarray}
Finally the transmission phase is given by
\begin{equation}
 \varphi_{t}=\tan^{-1}\left[ \frac{\chi^{+}\cos(\phi_{1}+(p_{x1}+p_{x5}))-\chi^{-}\sin(\phi_{1}-(p_{x1}+p_{x5}))}
 {\chi^{+}\sin(\phi_{1}-(p_{x1}+p_{x5}))+\chi^{-}\cos(\phi_{1}+(p_{x1}+p_{x5}))}\right].
\end{equation}

\begin{figure}[H]
\centering
\includegraphics[width=8cm, height=5cm]{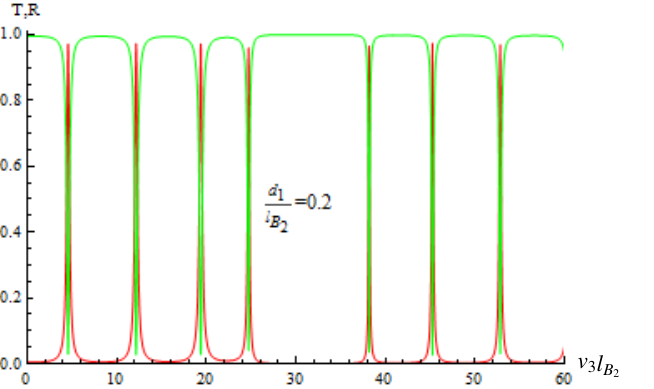}\ \ \ \ \ \ \ \ \ \ \ \
\includegraphics[width=5cm, height=7cm]{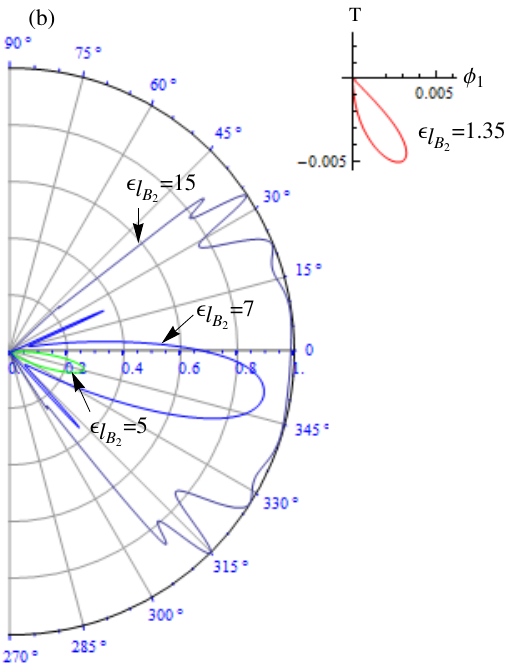}\\
%\end{figure}
%\begin{center}
 \caption{\sf{(a): Graphs depicting the reflection $R$ (green line) and transmission
$T$ (red line) coefficients as function of energy potential
$v_{3}l_{B_{3}}$ for the monolayer graphene barriers with
$\frac{d_{1}}{l_{B_{2}}}=0.2$, $\frac{d_{2}}{l_{B_{2}}}=0.8$,
$v_{2}l_{B_{2}} =26$, $\epsilon l_{B_{2}}=30$,
 $k_{y}l_{B_{2}}=1$, $\frac{l_{B_{3}}}{l_{B_{2}}}=2$ and $\mu l_{B_{2}}=4$. (b):
 Polar plot of a curve with radius (transmission $T$) as a
function of angle $\phi_{1}$ with  $\epsilon l_{B_{2}}=\{15, 7,
5,1.35\}$, $\frac{d_{1}}{l_{B_{2}}}=0.2$,
$\frac{d_{2}}{l_{B_{2}}}=0.8$, $\frac{l_{B_{3}}}{l_{B_{2}}}=0.6$,
 $v_{2}l_{B_{2}} =3.1$,
 $v_{3}l_{B_{2}}=1.2$ and $\mu l_{B_{2}}=2$.}}\label{fig2}
\end{figure}

Now we are ready for the computation of the 
transmission $T$ and reflection $R$ coefficients. For this purpose, we introduce the
associated current density $J$, which defines  $T$ and $R$ as
\begin{equation}
  T=\frac{ J_{\sf {tra}}}{ J_{\sf {inc}}},\qquad R=\frac{J_{\sf {ref}}}{ J_{\sf {inc}}}
\end{equation}
where $J_{\sf {\sf {inc}}}$, $J_{\sf {ref}}$ and $J_{\sf {\sf {tra}}}$ stand for the incident, reflected and transmitted components of
the current density, respectively. It is easy to show that the
current density $J$ reads as
\begin{equation}
J= e\upsilon_{F}\psi^{\dagger}\sigma _{x}\psi
\end{equation}
which gives the following results for the incident, reflected and
transmitted components
\bqr
J_{\sf {inc}} &=&  e\upsilon_{F}(\psi_{1}^{+})^{\dagger}\sigma
_{x}\psi_{1}^{+}
\\
  J_{\sf {ref}} &=& e\upsilon_{F} (\psi_{1}^{-})^{\dagger}\sigma _{x}\psi_{1}^{-}
\\
 J_{\sf {tra}}&=& e\upsilon_{F}(\psi_{5}^{+})^{\dagger}\sigma _{x}\psi_{5}^{+}.
\eqr
The energy conservation
\begin{equation}
\left[p_{x1}^{2}+(p_{y}+
 \frac{1}{l_{B_{2}}^{2}}(d_{1}-d_{2})-\frac{1}{l_{B_{3}}^{2}}d_{1})^{2}\right]^{\frac{1}{2}}=\left[p_{x5}^{2}+
  (p_{y}+\frac{1}{l_{B_{2}}^{2}}(d_{2}-d_{1})+\frac{1}{l_{B_{3}}^{2}}d_{1})^{2}\right]^{\frac{1}{2}}
\end{equation}
allows us to express the transmission and reflection
probabilities in the following simple forms
\begin{equation}
  T= \frac{p_{x5}}{p_{x1}}\frac{1}{|m_{11}|^{2}}, \qquad
  R=\left|\frac{m_{21}}{m_{11}}\right|^{2}.
\end{equation}
More explicitly the transmission coefficient $T$
reads as
\begin{equation}
T=\frac{4 p_{x5}(\cos\phi_{1})^{2}}{ p_{x1}
\left[(\chi^{+})^{2}+(\chi^{-})^{2}\right]}a_{n_{2}}^{4}b_{n_{2}}^{4}a_{n_{3}}^{2}
b_{n_{3}}^{2}\lambda_{n_{2}}^{2}\lambda_{n_{3}}^{2}\lambda_{n_{4}}^{2}.
\end{equation}
Obviously, $R$ and $T$ are not independent, they are related through
the unitarity requirement $T + R = 1$ that is clearly shown in Figure
\ref{fig2}a. Note that  \eqref{eq 51} implies that for certain
incidence angles $\phi_{1}$ the transmission is not allowed. In
fact for
\begin{equation}
\epsilon
l_{B_{2}}\leq\frac{1}{l_{B_{2}}}(d_{2}-d_{1})+\frac{d_{1}}{l_{B_{2}}}\left(\frac{l_{B_{2}}}{l_{B_{3}}}\right)^{2}
\end{equation}
all waves are completely reflected.

%In Figure \ref{fig2}
We show the numerical results for
the transmission, reflection coefficients and the
%Goos-H\"anchen like
GHL
shifts %, which are shown
in Figures
\ref{fig2}, \ref{figtt}, \ref{fig12}, \ref{figsr}, \ref{figi},
for several parameter values ($\epsilon$, $v_{2}$,
$v_{3}$, $\mu_j$ , $d_{1}$, $d_{2}$). For instance a typical value
of the magnetic field, say $B_{2} = 4T$, the magnetic length is
$l_{B_{2}} = 13nm$, and $\epsilon l_{B_{2}} = 1$ corresponding to
the energy $E = 44 meV$ \cite{martino}, these typical values will
serve to normalize the various variables.
The polar graph, Figure \ref{fig2}b,  shows the transmission as
a function of the incidence angle, the outermost circle
corresponds to full transmission, $T = 1$, while the origin of
this plot represents zero transmission. Requiring that
 $\epsilon l_{B_{2}}=3.7$, $d_{2}=d_{1}$, $l_{B_{2}}=l_{B_{3}}$, $v_{2}=v_{3}=0$, $\frac{d_{1}}{l_{B_{2}}}=\{0.5, 1.5, 3, 3.67\}$ and
 $\mu_j=0$ reproduces exactly the result obtained in previous work \cite{martino}. Similarly, the transmission as a function of energy $\epsilon$ for fixed
$\frac{d_{2}}{l_{B_{2}}}=0.8$, $\frac{d_{1}}{l_{B_{2}}}=0.2$ and
$\frac{l_{B_{3}}}{l_{B_{2}}}=0.6$, i.e.
$\frac{d_{2}-d_{1}}{l_{B_{2}}}+\frac{d_{1}}{l_{B_{2}}}\left(\frac{l_{B_{2}}}{l_{B_{3}}}\right)^{2}=1.156$,
shows that the transmission vanishes for $\epsilon l_{B_{2}}\leq 1.156$.

%\end{figure}

%%%%%%%%%%%%%%%%%%%%%%%%%%%%%%%%%%%%%%%%%%%%%%%%%%%%%%%
\section{GHL shifts for double barriers}
%%%%%%%%%%%%%%%%%%%%%%%%%%%%%%%%%%%%%%%%%%%%%%%%%%%%%%%%%
%\hspace{.33in}
In this section, we shall turn to the Goos-H\"anchen  like (GHL) shifts in
graphene by considering an incident, reflected and
transmitted beams around a given transverse wave vector $k_y =
k_{y_0}$ and angle of incidence $\phi_{1}(k_{y_{0}})\in [0,
\frac{\pi}{2}]$, denoted by the subscript $0$. These can be
expressed in integral form
\begin{eqnarray}
   \Psi_{i}(x,y) &=& \int_{-\infty}^{+\infty}dk_y\ f(k_y-k_{y_0})\ e^{i(k_{x1}(k_y)x+k_yy)}\left(
            \begin{array}{c}
              {1} \\
              {e^{i\phi_{1}(k_{y})}}
            \end{array}
          \right)\label{eq 79}\\
%\end{eqnarray}
%while the reflected wave is expressed as
%\begin{eqnarray}
\Psi_{r}(x,y) &=& \int_{-\infty}^{+\infty}dk_y\ r(k_y)\
f(k_y-k_{y_0})\ e^{i(-k_{x1}(k_y)x+k_yy)}\left(
            \begin{array}{c}
              {1} \\
              {-e^{-i\phi_{1}(k_{y})}} \\
            \end{array}
          \right)\label{refl}.
\end{eqnarray}
The reflection amplitude can be written as $r(k_y)=|r|e^{i\varphi_{r}}$ because of the $x$-component
of wavevector $k_{x1}$ as well as $\phi_{1}$ are function of $k_{y}$, where
each spinor plane wave is a solution of \eqref{eq 4}. The angular spectral distribution
$f(k_y-k_{y_0})$ can be assumed of Gaussian shape
\begin{equation}
f(k_y-k_{y_0})=w_ye^{-w_{y}^2(k_y-k_{y_0})^2}
\end{equation}
where $w_y$ being the half beam width at waist
\cite{Beenakker}. We can approximate the $k_{y}$-dependent terms
by a Taylor expansion around $k_{y_0}$ and retain only the first
order term to obtain
\begin{eqnarray}
&& \phi_{1}(k_{y})\approx
\phi_{1}(k_{y_{0}})+\frac{\partial\phi_{1}}{\partial
k_{y}}\Big|_{k_{y_{0}}}(k_{y}-k_{y_{0}})
\\
&& k_{x1}(k_{y})\approx k_{x1}(k_{y_{0}})+\frac{\partial
k_{x1}}{\partial k_{y}}\Big|_{k_{y_{0}}}(k_{y}-k_{y_{0}}).
\end{eqnarray}
The transmitted wave takes the form
\begin{eqnarray}
\Psi_{t}(x,y) &=& \int_{-\infty}^{+\infty}dk_y\ t(k_y)\
f(k_y-k_{y_0})\ e^{i(k_{x5}(k_y)x+k_yy)}\left(
            \begin{array}{c}
              {1} \\
              {e^{i\phi_{5}(k_{y})}} \\
            \end{array}
          \right)\label{trans}
\end{eqnarray}
where the transmission amplitude $t(k_y)=|t|e^{i\varphi_{t}}$ is calculated
through the use of boundary conditions. In order to determine the
GHL shifts of the transmitted beam through the graphene double
barriers, we adopt the following definition
 \cite{Chen1,Chen2}
 \begin{equation} \label{eq 15}
        S_{t}=- \frac{\partial \varphi_{t}}{\partial k_{y}}\Big|_{k_{y0}}.
 \end{equation}

\begin{figure}[h!]
\centering
\includegraphics[width=8cm, height=5cm]{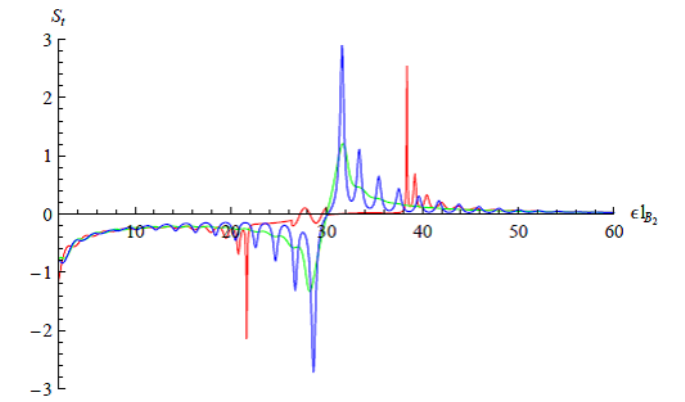}\\
\ \includegraphics[width=8cm, height=5cm]{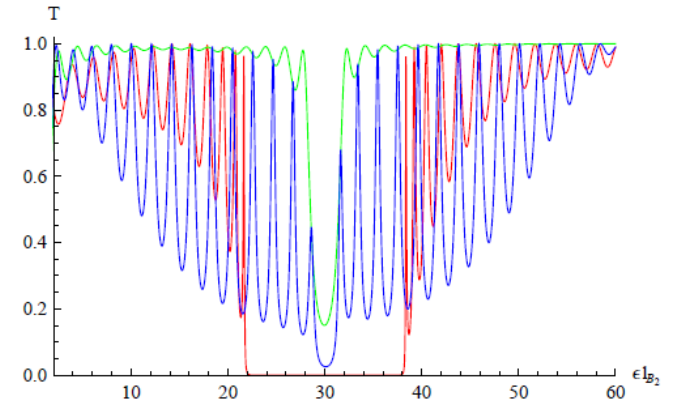}\\
%\end{figure}
%\begin{center}
 \caption{\sf{ The GHL shifts and the transmission as function of
energy $\epsilon l_{B_{2}}$ for the monolayer graphene
barriers with  $\frac{d_{2}}{l_{B_{2}}}=0.8$, $v_{2} l_{B_{2}}
=30$, $v_{3} l_{B_{2}}=30$,
 $k_{y} l_{B_{2}}=1$, $\frac{l_{B_{2}}}{l_{B_{3}}}=0.5$,
 $\frac{d_{1}}{l_{B_{2}}}=0.7$ where ($\mu_{2}l_{B_{2}}=0$, $\mu_{3}l_{B_{2}}=8$) color
 red, ($\mu_{2}l_{B_{2}}=0$, $\mu_{3}l_{B_{2}}=0$) color
 green and ($\mu_{2}l_{B_{2}}=8$, $\mu_{3}l_{B_{2}}=0$)  color blue.}}\label{figtt}
\end{figure}
%\end{figure}
In Figure \ref{figtt}, the above transmission and GHL shifts are
shown versus energy $\epsilon l_{B_{2}}$ for different
parameters of our system
$\left(\frac{d_{2}}{l_{B_{2}}}=0.8, v_{2} l_{B_{2}} =30, v_{3}
l_{B_{2}}=30,
 k_{y} l_{B_{2}}=1, \frac{l_{B_{3}}}{l_{B_{2}}}=2,
 \frac{d_{1}}{l_{B_{2}}}=0.7\right)$ with zero-gap $\left(\mu_{2}l_{B_{2}}=\mu_{3}l_{B_{2}}=0\right)$: green color and finite gap
($\mu_{2}l_{B_{2}}=0$,
 $\mu_{3}l_{B_{2}}$=8): red color and  ($\mu_{2}l_{B_{2}}=8$,
 $\mu_{3}l_{B_{2}}$=0): blue color. It is clearly seen that GHL shifts are oscillating
between negative and positive values around the critical point
$\epsilon l_{B} = v_{2}l_{B} = v_{3}l_{B}$. The quantity $k_{y}
l_{B} = m^{\ast}$ plays a very important role in the transmission
of Dirac fermions via obstacles created by a series of
scattering potentials, because it is associated with the effective
mass of the particle and hence determines the threshold for
allowed energies. However, in the presence of an inhomogeneous
magnetic field in the regions $|x|\leq d_{2}$, it reduces this
effective mass to
$\left(p_{y}+\frac{1}{l_{B_{2}}^{2}}\left(d_{1}-d_{2}\right)-\frac{d_1}{l_{B_{3}}^{2}} \right)$
in the incidence region while it increases it to
$\left(p_{y}-\frac{1}{l_{B_{2}}^{2}}\left(d_{1}-d_{2}\right)+\frac{d_1}{l_{B_{3}}^{2}}\right)$
in the transmission region. The allowed energies are then
determined by the greater effective mass condition:
$$\epsilon
l_{B_{2}} \geq
p_{y}-\frac{1}{l_{B_{2}}^{2}}(d_{1}-d_{2})+\frac{d_1}{l_{B_{3}}^{2}}$$.\\

\begin{figure}[h!]
\centering
\ \includegraphics[width=8cm, height=5cm]{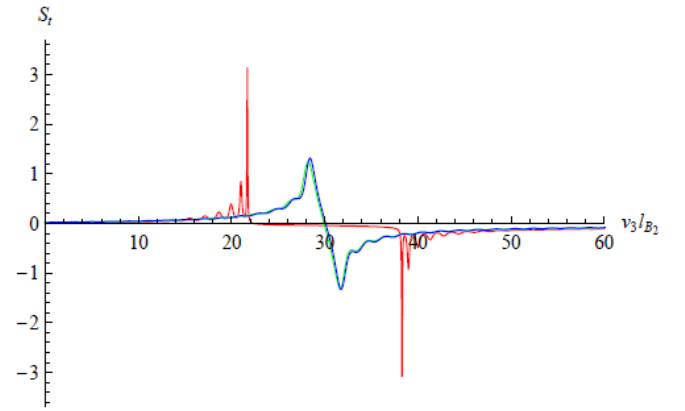}\\
\includegraphics[width=8cm, height=5cm]{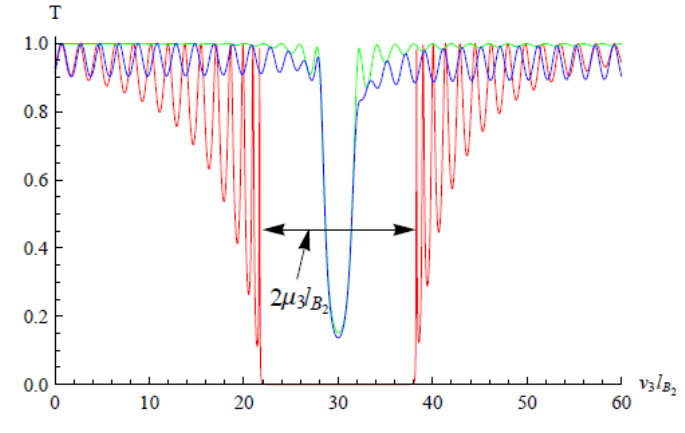}\\
%\end{figure}
%\begin{center}
 \caption{\sf{The GHL shifts and the transmission as function of
the potential $v_{3} l_{B_{2}}$ for the monolayer graphene
barriers with  $\frac{d_{2}}{l_{B_{2}}}=0.8$, $v_{2} l_{B_{2}}
=27$, $\epsilon l_{B_{2}}=30$,
 $k_{y} l_{B_{2}}=1$, $\frac{l_{B_{3}}}{l_{B_{2}}}=2$,
 $\frac{d_{1}}{l_{B_{2}}}=0.78$ where ($\mu_{2}l_{B_{2}}=0$, $\mu_{3}l_{B_{2}}=8$)  color
 red, ($\mu_{2}l_{B_{2}}=0$, $\mu_{3}l_{B_{2}}=0$) color
 green and ($\mu_{2}l_{B_{2}}=8$, $\mu_{3}l_{B_{2}}=0$) color blue.}}\label{fig12}
\end{figure}
%\end{figure}
The above GHL shifts and transmission are plotted in Figure
\ref{fig12} in terms of the potential $v_{3} l_{B_{2}}$ for some values of the physical parameters. It is clearly seen that $S_t$ is oscillating between
negative and positive values around the critical point
$v_{3}l_{B_{2}} =\epsilon l_{B_{2}}$. At such points the transmission  vanishes for $\epsilon l_{B_{2}}-\mu_{3} l_{B_{2}}\leq
v_{3}l_{B_{2}}\leq \epsilon l_{B_{2}}+\mu_{3} l_{B_{2}} $ and
oscillates otherwise.\\

\begin{figure}[h!]
\centering
\includegraphics[width=8cm, height=5cm]{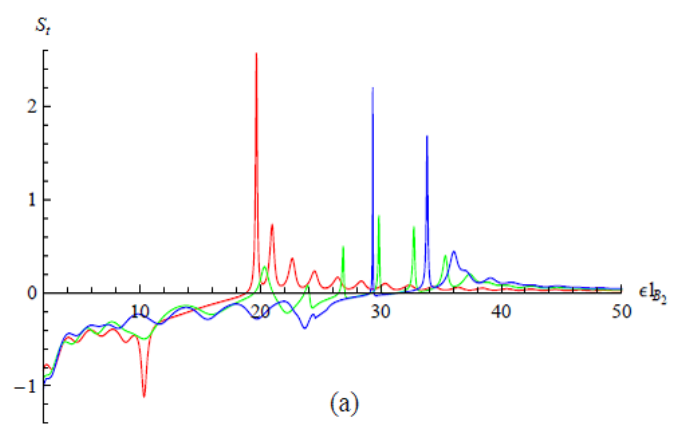}\ \ \ \
\includegraphics[width=8cm, height=5cm]{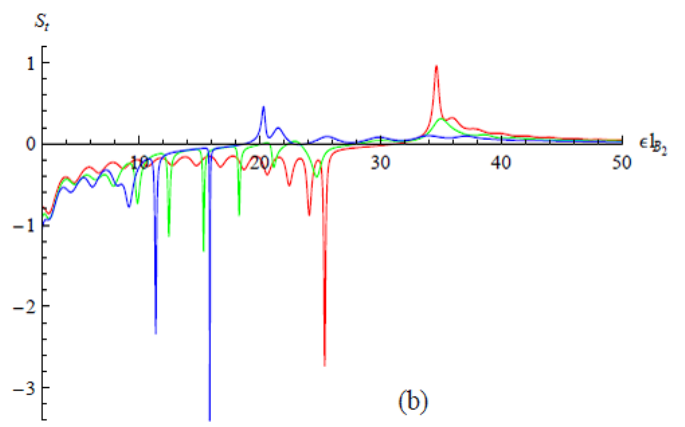}\\
\includegraphics[width=8cm, height=5cm]{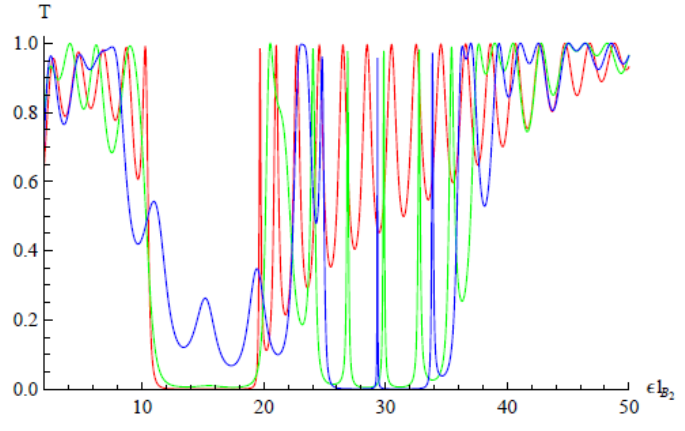}\ \ \ \
\includegraphics[width=8cm, height=5cm]{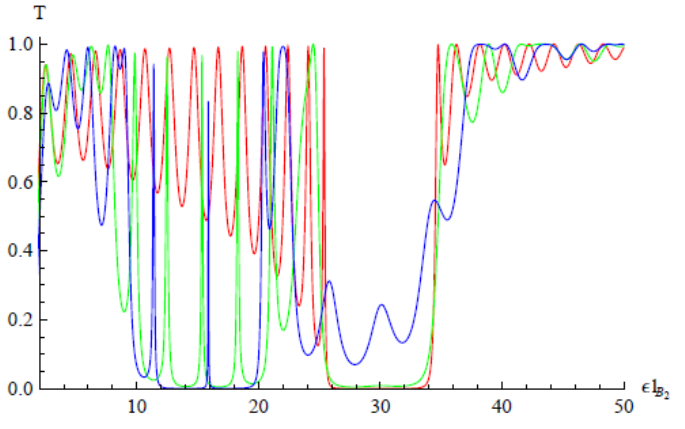}\\
%\end{figure}
%\begin{center}
 \caption{\sf{The GHL shifts and the transmission as function of
energy $\epsilon l_{B_{2}}$ for the monolayer graphene barriers.
(a)/(b)  with ($v_{2}l_{B_{2}}=30$,
$v_{3}l_{B_{2}}=15$)/($v_{2}l_{B_{2}}=15$, $v_{3}l_{B_{2}}=30$),
with $\frac{d_{2}}{l_{B_{2}}}=0.8$,
 $k_{y} l_{B_{2}}=1$, $\frac{l_{B_{3}}}{l_{B_{2}}}=2$, $\mu_{2}
 l_{B_{2}}=\mu_{3}
 l_{B_{2}}=4$,
 $\frac{d_{1}}{l_{B_{2}}}=0.19$ (blue line), $\frac{d_{1}}{l_{B_{2}}}=0.4$ (green line) and $\frac{d_{1}}{l_{B_{2}}}=0.7$ (red line).}}\label{figsr}
\end{figure}
%\end{figure}
In Figure \ref{figsr}, the transmission and GHL shifts are
shown versus energy $\epsilon l_{B_{2}}$. One can notice
that at the Dirac points ($\epsilon l_{B_{2}}= v_{2}l_{B_{2}}$,
$\epsilon l_{B_{2}} = v_{3}l_{B_{2}}$), the GHL shifts change their
sign. This change %in sign of
%the GH shifts
shows clearly that they are strongly dependent on
the barrier heights. We also observe that the GHL shifts are
positive as long as the energy satisfies the condition $\epsilon
l_{B_{2}}>v_{2}l_{B_{2}}>v_{3}l_{B_{2}}$ (Figure
\ref{figsr}a) and negative for
$\epsilon l_{B_{2}}<v_{2}l_{B_{3}}<v_{2}l_{B_{2}}$ (Figure
\ref{figsr}b.)
%. In . the GH shifts are positive as long as the energy
%satisfies the condition $\epsilon
%l_{B_{2}}>v_{3}l_{B_{2}}>v_{2}l_{B_{2}}$ and negative for
%$\epsilon l_{B_{2}}<v_{3}l_{B_{2}}<v_{2}l_{B_{2}}$

\begin{figure}[H]
\centering
\includegraphics[width=8cm, height=5cm]{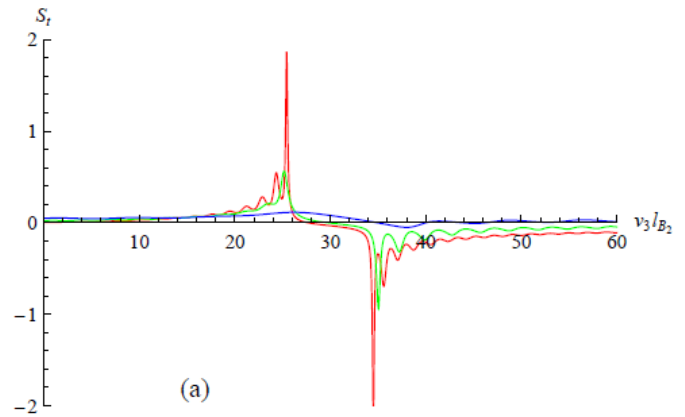}
\includegraphics[width=8cm, height=5cm]{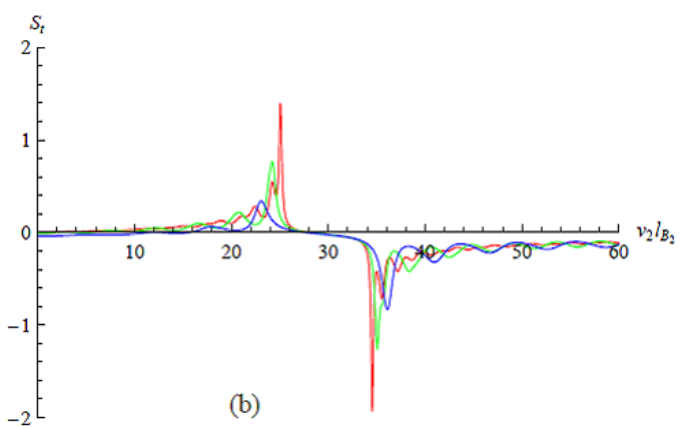}\\
\includegraphics[width=8cm, height=5cm]{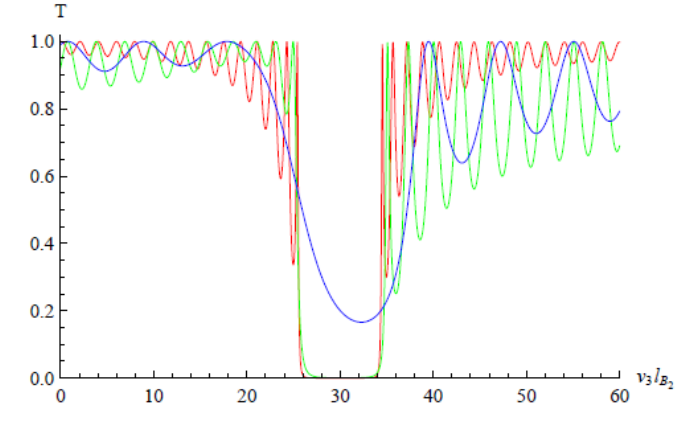}\ \
\includegraphics[width=8cm, height=5cm]{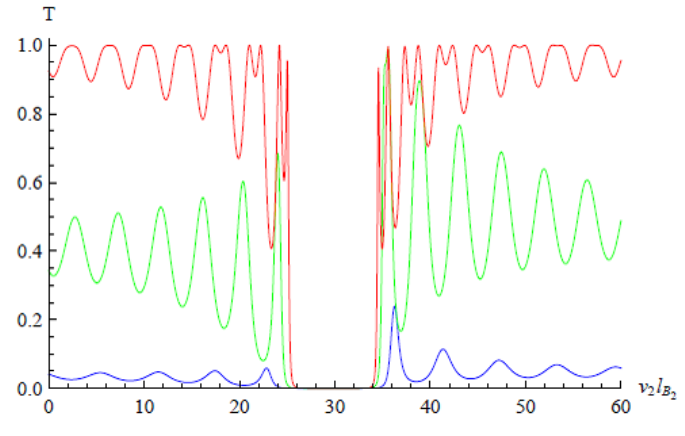}  \ \ \ \\
%\end{figure}
%\begin{center}
 \caption{\sf{The GHL shifts and the transmission as function of
energy potential $v_{3}l_{B_{2}}$ and $v_{2}l_{B_{2}}$ for the
monolayer graphene barriers. (a): $\epsilon l_{B_{2}}=30$,
$v_{2}l_{B_{2}}=15$, $\frac{d_{2}}{l_{B_{2}}}=0.8$,
 $k_{y} l_{B_{2}}=1$, $\frac{l_{B_{3}}}{l_{B_{2}}}=2$, $\mu_{2}
 l_{B_{2}}=\mu_{3}
 l_{B_{2}}=4$,
 $\frac{d_{1}}{l_{B_{2}}}=0.2$ (blue line), $\frac{d_{1}}{l_{B_{2}}}=0.5$ (green line) and $\frac{d_{1}}{l_{B_{2}}}=0.78$ (red line). (b):
 $\epsilon l_{B_{2}}=30$,
$v_{3}l_{B_{2}}=32$, $\frac{d_{2}}{l_{B_{2}}}=0.8$,
 $k_{y} l_{B_{2}}=1$, $\frac{l_{B_{3}}}{l_{B_{2}}}=2$, $\mu
 l_{B_{2}}=4$,
 $\frac{d_{1}}{l_{B_{2}}}=0.3$ (blue line), $\frac{d_{1}}{l_{B_{2}}}=0.12$ (green line) and $\frac{d_{1}}{l_{B_{2}}}=0.02$ (red line).}}\label{figi}
\end{figure}

In Figure \ref{figi}, we analyze the transmission
coefficients versus the potential $v_{3}l_{B_{2}}$ and
$v_{2}l_{B_{2}}$. In doing so, we fix the energy $\epsilon
l_{B_{2}}=30$ and choose a value of $\frac{d_{1}}{l_{B_{2}}}$,
then we compute the transmission as shown in Figure \ref{figi}a.
We notice that the transmission decreases if $\frac{d_{1}}{l_{B_{2}}}$ increases
and then vanishes while Figure \ref{figi}b shows different behavior.
Note that, the Dirac points represent the zero modes for Dirac operator
\cite{Sharma19} and lead to the emergence of new Dirac points, which
has been discussed in different works \cite{Bhattacharjee,Park2}.
Such points separate the two regions of positive and negative
refraction. In cases where $v_{2}l_{B_{2}} < \epsilon l_{B_{2}}$
and $v_{2}l_{B_{2}}  > \epsilon l_{B_{2}}$ (respectively
$v_{3}l_{B_{2}} < \epsilon l_{B_{2}}$  and $v_{2}l_{B_{2}} >
\epsilon l_ {B_{2}}$), the shifts are respectively in the forward
and backward directions, due to the fact that the signs of the group
velocity are opposite. \\

%\end{figure}

%%%%%%%%%%%%%%%%%%%%%%%%%%%%%%%%%%%%%%%%%%%%%%%%%%%
\section{Conclusion}
%%%%%%%%%%%%%%%%%%%%%%%%%%%%%%%%%%%%%%%%%%%%%%%
To conclude, we have studied the transport of electrons in
graphene scattered by double barrier in the presence of an
inhomogeneous magnetic field. We obtained the solutions
for the energy spectrum taking into account the conservation energy and
noticed that for certain incidence angles the
transmission is not allowed for $\epsilon
l_{B_{2}} \leq
p_{y}-\frac{1}{l_{B_{2}}^{2}}(d_{1}-d_{2})+\frac{1}{l_{B_{3}}^{2}}d_{1}$.
However, the transmission probability $T$ does not vanish in
general, we also found that, in contrast to electrostatic barriers,
magnetic barriers are able to confine Dirac fermions. This allowed us
to calculate the GHL shifts of reflected and transmitted electron
beams in a graphene double barrier structure in the presence of an inhomogeneous magnetic
field. We also established some correlation between the electronic transport
properties of Dirac fermions with the GHL shifts.

The numerical data showed how these shifts behave in relation to the transmission probability $T$.
It is found that the GHL shifts can be modulated by the
incident energy $\epsilon l_{B_{2}}$, potential energies
$v_{2}l_{B_{2}}$ and  $v_{3}l_{B_{2}}$. The GHL shifts
still change sign, but the point where it changes sign has been displaced
to the left and the absolute value of the maximum of the shifts
increased as well. Thus we seen that the GHL shifts in the transmission region
can be either negative or positive.

%%%%%%%%%%%%%%%%%%%%%%%%%%%%%%%%%%%%%%%%%%%%%%%%%%%
\section*{Acknowledgments}
%%%%%%%%%%%%%%%%%%%%%%%%%%%%%%%%%%%%%%%%%%%%%%%

The generous support provided by the Saudi Center
for Theoretical Physics (SCTP) is highly appreciated by all
authors. H.B. Also acknowledges the support of King Fahd University of Petroleum and Minerals
to the Theoretical Physics Research Group.

\end{document}